%% file: main.tex
\documentclass[runningheads]{new_tlp}
\usepackage[T1]{fontenc}
\usepackage{graphicx}
\usepackage{algorithm}
\usepackage[noend]{algpseudocode}
\usepackage{amsmath, amsfonts}
\usepackage{bussproofs}
\usepackage{fancyvrb}
\usepackage{listings}
\usepackage{mathtools}
\usepackage{tikz-cd}
\usetikzlibrary{shapes.geometric,shapes,positioning,calc,patterns,shapes.arrows,decorations.pathreplacing}
\usepackage{subcaption}
\usepackage{hyperref}
\usepackage{pifont}  

\newcommand{\EDB}{\mathrm{EDB}}
\newcommand{\IDB}{\mathrm{IDB}}
\newcommand{\dlimpl}{\ \texttt{:-}\ }

\newcommand{\brackets}[1]{({#1})}

\DeclareMathOperator{\prov}{inc-prov}
\DeclareMathOperator{\allprov}{all-inc-prov}

\newtheorem{definition}{Definition}[section]
 
\definecolor{dkgreen}{rgb}{0,0.6,0}
\definecolor{gray}{rgb}{0.5,0.5,0.5}
\definecolor{mauve}{rgb}{0.58,0,0.82}

\lstdefinestyle {DatalogStyle} {frame=tb,
  language=C,
  aboveskip=3mm,
  belowskip=3mm,
  escapechar=`,
  showstringspaces=false,
  columns=flexible,
  basicstyle={\footnotesize\ttfamily},
  numbers=none,
  numberstyle=\tiny\color{gray},
  commentstyle=\color{dkgreen},
  stringstyle=\color{mauve},
  breaklines=true,
  breakatwhitespace=true,
  tabsize=3,
  linewidth=\linewidth
}

\lstdefinestyle {C++Style} {frame=tb,
  language=C++,
  aboveskip=3mm,
  belowskip=3mm,
  escapechar=`,
  showstringspaces=false,
  columns=flexible,
  basicstyle={\footnotesize\ttfamily},
  numbers=left,
  numberstyle=\tiny\color{gray},
  keywordstyle=\color{blue},
  commentstyle=\color{dkgreen},
  stringstyle=\color{mauve},
  breaklines=true,
  breakatwhitespace=true,
  tabsize=3,
  linewidth=\linewidth
}
\title{Provenance Guided Rollback Suggestions} 

\author[D. Zhao et al.]
    {DAVID ZHAO\\
        University of Sydney
    \and PAVLE SUBOTI\'C\\
        Microsoft
    \and MUKUND RAGHOTHAMAN\\
        University Southern California
    \and BERNHARD SCHOLZ\\
        University of Sydney}

\begin{document}

\maketitle

\input{src/abstract.tex}


\input{src/intro.tex}
\input{src/background.tex}
\input{src/inc-prov.tex}
\input{src/fault-localization.tex}
\input{src/impl.tex}
\input{src/exp.tex}
\input{src/related.tex}
\input{src/concl.tex}

\subsubsection*{Acknowledgments}

M.R. was funded by U.S. NSF grants CCF-2146518, CCF-2124431, and CCF-2107261.

\bibliography{references}

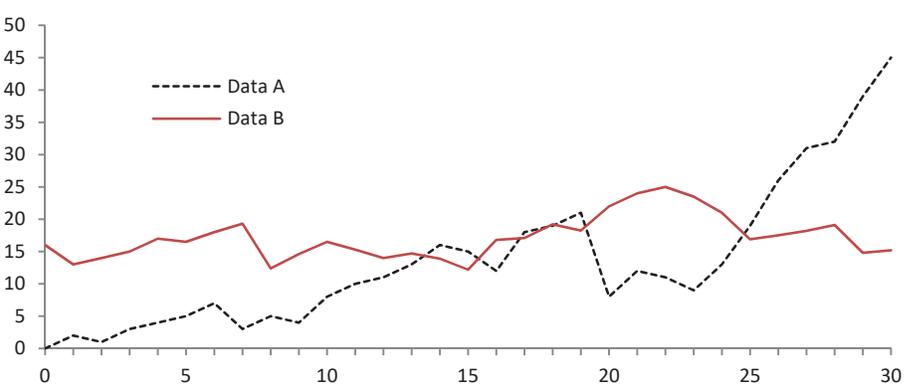
\end{document}

%% file: src/abstract.tex
\begin{abstract}
    Advances in incremental Datalog evaluation strategies have made Datalog
    popular among use cases with constantly evolving inputs such as static
    analysis in continuous integration and deployment pipelines. As a result,
    new logic programming debugging techniques are needed to support these
    emerging use cases. 

    This paper introduces an incremental debugging technique for Datalog, which
    determines the failing changes for a \emph{rollback} in an incremental
    setup. Our debugging technique leverages a novel incremental provenance
    method. We have implemented our technique using an incremental version of
    the Souffl\'{e} Datalog engine and evaluated its effectiveness on the DaCapo
    Java program benchmarks analyzed by the Doop static analysis library.
    Compared to state-of-the-art techniques, we can localize faults and suggest
    rollbacks with an overall speedup of over 26.9$\times$ while providing
    higher quality results.
\end{abstract}

%% file: src/intro.tex
\section{Introduction}
Datalog has achieved widespread adoption in recent years, particularly in static
analysis use cases~\cite{doop09,pointsto15,mm18,icse19,eq10,de11,tiros} that can
benefit from incremental evaluation. In an industrial setting, static analysis
tools are deployed in continuous integration and deployment setups to perform
checks and validations after changes are made to a code
base~\cite{infercicd,codeql}. Assuming that changes between analysis runs
(aka.~epochs) are small enough, a static analyzer written in Datalog can be
effectively processed by incremental evaluation
strategies~\cite{zhao2021towards,motik2019maintenance,ryzhyk2019differential,mcsherry2013differential}
which recycle computations of previous runs. When a fault appears from a change
in the program, users commonly need to (1) localize which changes caused the
fault and (2) partially roll back the changes so that the faults no longer
appear. However, manually performing this bug localization and the subsequent
rollback is impractical, and users typically perform a full rollback while
investigating the fault's actual cause~\cite{reverts,backtracking}. The correct
change is re-introduced when the fault is found and addressed, and the program
is re-analyzed. This entire debugging process can take significant time. Thus,
an automated approach for detecting and performing partial rollbacks can
significantly enhance developer productivity.

\begin{figure}[h]
    \centering
    \begin{tikzpicture}
        \node[] (E1) at (-0.2,2) {$\text{Source Program}_1$};
        \node[shape=rectangle, draw=black, minimum width=0.8cm, rounded corners=0.1cm] (P1) at (3,1) {$\Delta P$};


        \node[] (E2) at (-0.2,0) {$\text{Source Program}_2$};

        \path[->] (E1) edge (E2);
        \node[] (delta) at (0.3,1) {$\Delta E$};

        \path[->] (delta) edge (P1);

        \node[] (I) at (5.6,1) {$\Delta I$};
        \path[->] (P1) edge (I);

        \node[anchor=center] at (7.2,2.38) {\scriptsize Unchanged Tuples};
        \draw[fill=black!10] (6.4,2.2) rectangle (8,2.0);
        \draw[fill=black!10] (6.4,2.0) rectangle (8,1.8);
        \node[anchor=center] at (7.2,1.58) {\scriptsize Inserted};
        \draw[fill=green!20] (6.4,1.4) rectangle (8,1.2);
        \draw[fill=green!20] (6.4,1.2) rectangle (8,1.0);
        \draw[fill=red!20] (6.4,1.0) rectangle (8,0.8);
        \node[anchor=center] at (7.2,0.58) {\scriptsize Deleted};
        \draw[fill=green!20] (6.4,0.4) rectangle (8,0.2);
        \draw[fill=green!20] (6.4,0.2) rectangle (8,0.0);
        \draw[fill=red!20] (6.4,0.0) rectangle (8,-0.2);

        \draw[decorate,decoration={brace,aspect=0.64}] (6.1,-0.2) -- (6.1,1.65);




        \node[anchor=center] at (6.3,0.9) {\tiny \ding{55}};
        \node[anchor=center] at (6.3,-0.1) {\tiny \ding{55}};

        \node[anchor=center] at (8.2,0.9) {\tiny $t_1$};
        \node[anchor=center] at (8.2,-0.1) {\tiny $t_2$};

    \end{tikzpicture}
    \caption{A scenario where an incremental update results in faults in the
    output}
    \label{fig:incprov:scenario}
\end{figure}
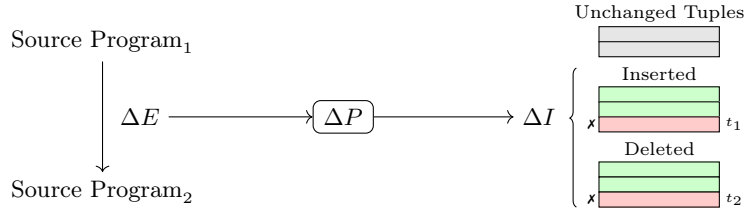

For instance, consider the example in Figure~\ref{fig:incprov:scenario}. The
diagram shows the use of incremental evaluation for program analysis use cases.
On the left, the source program is updated, resulting in a change $\Delta E$,
which is input to the incremental program analysis $\Delta P$. After computing
the incremental update, some result tuples are unchanged, some are inserted, and
some are deleted. However, some of the changes (insertions or deletions) may be
\emph{unwanted} (i.e., the user does not agree with the change), and hence we
can view these as \emph{faults} that appeared as a result of the incremental
update.


Existing state-of-the-art Datalog debugging techniques that are available employ
data provenance~\cite{provgraph,zhao2020debugging} or algorithmic
debugging~\cite{caballero2017survey} to provide explanations. However, these
techniques require a deep understanding of the tool's implementation and target
the ruleset, not the input. Therefore, such approaches are difficult to apply to
automate input localization and rollback. The most natural candidate for this
task is \emph{delta debugging}~\cite{zeller1999yesterday,zeller2002simplifying},
a debugging framework for generalizing and simplifying a failing test case. This
technique has recently been shown to scale well when integrated with
state-of-the-art Datalog synthesizers~\cite{raghothaman2019provenance} to obtain
better synthesis constraints. Delta debugging uses a divide-and-conquer approach
to localize the faults when changes are made to a program, thus providing a
concise witness for the fault. However, the standard delta debugging approach is
programming language agnostic and requires programs to be re-run, which may
require significant time. 

In this paper, we introduce a novel approach to automating localize-rollback
debugging. Our approach comprises a novel incremental provenance technique and
two intertwined algorithms that diagnose and compute a rollback suggestion for a
set of faults (missing and unwanted tuples). The first algorithm is a
\emph{fault localization} algorithm that reproduces a set of faults, aiding the
user in diagnosis. Fault localization traverses the incremental proof tree
provided by our provenance technique, producing the subset of an incremental
update that causes the faults to appear in the current epoch. The second
algorithm performs an \emph{input repair} to provide a local \emph{rollback
suggestion} to the user. A rollback suggestion is a subset of an incremental
update, such that the faults are fixed when it is rolled back.

We have implemented our technique using an extended incremental version of the
Souffl\'{e}~\cite{jordan2016souffle,zhao2021towards} Datalog engine and
evaluated its effectiveness on DaCapo~\cite{DaCapo} Java program benchmarks
analyzed by the Doop~\cite{doop09} static analysis tool. Compared to delta
debugging, we can localize and fix faults with a speedup of over 26.9$\times$
while providing smaller repairs in 27\% of the benchmarks. To the best of our
knowledge, we are the first to offer such a debugging feature in a Datalog
engine, particularly for large workloads within a practical amount of time. We
summarize our contributions as follows:

\begin{itemize}
    \item We propose a novel debugging technique for incremental changing input.
        We employ localization and rollback techniques for Datalog that scale
        to real-world program analysis problems. 
    \item We propose a novel incremental provenance mechanism for Datalog
        engines. Our provenance technique leverages incremental information to
        construct succinct proof trees.
    \item We implement our technique in the state-of-the-art Datalog engine
        Souffl\'{e}, including extending incremental evaluation to compute
        provenance. 
    \item We evaluate our technique with Doop static analysis for large Java
        programs and compare it to a delta-debugging approach adapted for the
        localization and rollback problem. 
\end{itemize}

%% file: src/background.tex
\section{Overview}
\label{sec:background}

\begin{figure}[htbp]
\vspace{-2em}
\begin{subfigure}[b]{0.5\textwidth}
    \begin{lstlisting}[style=C++Style]
admin = new Admin();
sec = new AdminSession();
ins = new InsecureSession();
admin.session = ins;
if (admin.isAdmin && admin.isAuth)
    admin.session = sec;
else
    userSession = ins;
    \end{lstlisting}
    \vspace{-1em}
    \caption{Input Program\label{ex:program}}
\end{subfigure}
\hspace{1mm}
\begin{subfigure}[b]{0.4\textwidth}
    \begin{lstlisting}[style=DatalogStyle]
new(admin,L1).
new(sec,L2).
new(ins,L3).
store(admin,session,ins).

store(admin,session,sec).

assign(userSession,ins). 
    \end{lstlisting}
    \vspace{-1em}
    \caption{EDB Tuples\label{ex:relation}}
\end{subfigure}

\vspace{2mm}
\begin{subfigure}[b]{0.927\textwidth}
    \begin{lstlisting}[style=DatalogStyle]
// r1: var = new Obj()
vpt(Var, Obj) :- new(Var, Obj).
// r2: var = var2
vpt(Var, Obj) :- assign(Var, Var2), vpt(Var2, Obj).
// r3: v = i.f; i2.f = v2 where i, i2 point to same obj
vpt(Var, Obj) :- load(Var, Inter, F), store(Inter2, F, Var2), 
                   vpt(Inter, InterObj), vpt(Inter2, InterObj), 
                   vpt(Var2, Obj).
// r4: v1, v2 point to same obj
alias(V1, V2) :- vpt(V1, Obj), vpt(V2, Obj), V1 != V2.
    \end{lstlisting}
    \vspace{-1em}
    \caption{Datalog Points-to Analysis\label{ex:analysis}}
\end{subfigure}
\caption{Program Analysis Datalog Setup~\label{fig:example}}
\vspace{-1em}
\end{figure}

\subsection{Motivating Example}
Consider a Datalog points-to analysis in Fig.~\ref{fig:example}. Here, we show an
input program to analyze (Fig.~\ref{ex:program}), which is encoded as a set of
tuples (Fig.~\ref{ex:relation}) by an \emph{extractor}, which maintains a
mapping between tuples and source
code~\cite{jordan2016souffle,Schfer2017AlgebraicDT,vallee2010soot}. We have
relations \texttt{new}, \texttt{store}, \texttt{load}, and \texttt{assign}
capturing the semantics of the input program to analyze. These relations are
also known as the \emph{Extensional Database} (EDB), representing the analyzer's
input. The analyzer written in Datalog computes relations \texttt{vpt}
(\emph{Variable Points To}) and \texttt{alias} as the output, which is also
known as the \emph{Intensional Database} (IDB). For the points-to analysis,
Fig.~\ref{ex:analysis} has four rules. A rule is of the form: $$R_h(X_h) \dlimpl
R_1(X_1), \ldots, R_k(X_k).$$ Each $R(X)$ is an \emph{atom}, with $R$ being a
\emph{relation} name and $X$ being a vector of \emph{variables} and
\emph{constants} of appropriate arity. The predicate to the left of $\dlimpl$ is
the \emph{head} and the sequence of predicates to the right is the \emph{body}.
A Datalog rule can be read from right to left: ``for all rule instantiations, if
every tuple in the body is derivable, then the corresponding tuple for the head
is also derivable''.

For example, $r_2$ is $\mathtt{vpt}(\mathtt{Var}, \mathtt{Obj}) \dlimpl
\mathtt{assign}(\mathtt{Var}, \mathtt{Var2}),\ \mathtt{vpt}(\mathtt{Var2},
\mathtt{Obj})$, which can be interpreted as ``if there is an assignment from
\texttt{Var} to \texttt{Var2}, and if \texttt{Var2} may point to \texttt{Obj},
then also \texttt{Var} may point to \texttt{Obj}''. In combination, the four
rules represent a \emph{flow-insensitive} but \emph{field-sensitive} points-to
analysis. The IDB relations \texttt{vpt} and \texttt{alias} represent the
analysis result: variables may point to objects and pairs of variables that may
be an alias with each other.

Suppose the input program in Fig.~\ref{ex:program} changes by adding a method to
upgrade a user session to an admin session with the code:
\begin{verbatim}
    upgradedSession = userSession;
    userSession = admin.session;
\end{verbatim}
The result of the points-to analysis can be incrementally updated by inserting
the tuples \texttt{assign(upgradedSession, userSession)} and
\texttt{load(userSession, admin, session)}. After computing the incremental
update, we would observe that \texttt{alias(userSession, sec)} is now contained
in the output. However, we may wish to maintain that \texttt{userSession}
\emph{should not} alias with the secure session \texttt{sec}. Consequently, the
incremental update has introduced a \emph{fault}, which we wish to localize and
initiate a rollback.

A fault localization provides a subset of the incremental update that is
sufficient to reproduce the fault, while a rollback suggestion is a subset of
the update which fixes the faults. In this particular situation, the fault
localization and rollback suggestion are identical, containing only the
insertion of the second tuple, \texttt{load(userSession, admin, session)}.
Notably, the other tuple in the update, \texttt{assign(upgradedSession,
userSession)}, is irrelevant for reproducing or fixing the fault and thus is not
included in the fault localization/rollback.

In general, an incremental update may contain thousands of inserted and deleted
tuples, and a set of faults may contain multiple tuples that are changed in the
incremental update. Moreover, the fault tuples may have multiple alternative
derivations, meaning that the localization and rollback results are different.
In these situations, automatically localizing and rolling back the faults to
find a small relevant subset of the incremental update is essential to provide a
concise explanation of the faults to the user.

The scenario presented above is common during software development, where making
changes to a program causes faults to appear. While our example concerns a
points-to analysis computed for a source program, our fault localization and
repair techniques are, in principle, applicable to any Datalog program.

\paragraph{\textbf{Problem Statement:}} Given an incremental update with its
resulting faults, automatically find a fault localization and rollback
suggestion.

\subsection{Approach Overview}

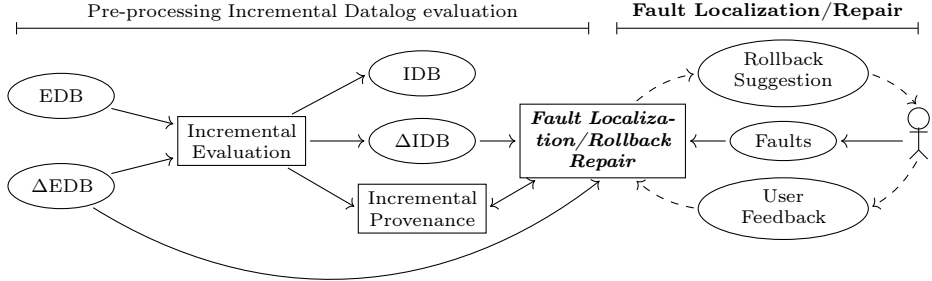
\begin{figure}
\centering
    \footnotesize
    \begin{tikzpicture}[scale=0.6, every node/.append style={font=\scriptsize}]
        \usetikzlibrary{shapes}
        \node [draw, ellipse, minimum height=0.6cm, align=center, text width=0.8cm, name=edb] at (0, 1) {EDB};
        \node [draw, ellipse, minimum height=0.6cm, align=center, text width=0.8cm, name=delta_edb] at (0, -1){$\Delta$EDB};

        \node [draw, rectangle, align=center, text width=1.5cm, name=eval] at (4, 0) {Incremental Evaluation};

        \node [draw, ellipse, minimum height=0.6cm,align=center, text width=0.8cm, name=idb] at (8, 1.5) {IDB};
        \node [draw, ellipse, minimum height=0.6cm,align=center, text width=0.8cm, name=delta_idb] at (8, 0) {$\Delta$IDB};

        \node [draw, rectangle, align=center, text width=1.5cm, name=prov] at (8, -1.5) {Incremental Provenance};

        \node [draw, rectangle, align=center, text width=2cm, name=localize] at (12, 0) {\textbf{\emph{Fault Localization/Rollback Repair}}};

        \node [draw, ellipse, align=center, text width=0.8cm, name=faults] at (16, 0) {Faults};
        \node [draw, ellipse, text height=0.1cm, align=center, text width=1.4cm, name=repair] at (16, 1.5) {Rollback Suggestion};
        \node [draw, ellipse, text height=0.1cm, align=center, text width=1.4cm, name=feedback] at (16, -1.5) {User Feedback};

        \node [draw, circle, minimum size=0.7, name=person_head] at (19, 0.5) {};
        \draw (19, 0.3) -- (19, -0.2);
        \draw (19.2, 0.2) -- (18.8, 0.2);
        \draw (19, -0.2) -- (18.8, -0.4);
        \draw (19, -0.2) -- (19.2, -0.4);

        \node [minimum size=0, name=person_middle] at (19, 0) {};
        \node [minimum size=0, name=person_feet] at (19, -0.2) {};

        \path [draw, ->, shorten >= 1.5pt, shorten <= 1.5pt] (edb) -- (eval);
        \path [draw, ->, shorten >= 1.5pt, shorten <= 1.5pt] (delta_edb) -- (eval);
        \path [draw, ->, shorten >= 1.5pt, shorten <= 1.5pt] (eval) -- (delta_idb);
        \path [draw, ->, shorten >= 1.5pt, shorten <= 1.5pt] (eval) -- (idb.west);
        \path [draw, ->, shorten >= 1.5pt, shorten <= 1.5pt] (eval) -- (prov.west);
        \path [draw, ->, shorten >= 1.5pt, shorten <= 1.5pt] (delta_idb.east) -- (localize);
        \path [draw, <->, shorten >= 1.5pt, shorten <= 1.5pt] (prov.east) -- (localize);

        \path [draw, ->, shorten >= 1.5pt, shorten <= 3pt] (person_middle) -- (faults.east);
        \path [draw, ->, shorten >= 1.5pt, shorten <= 1.5pt] (faults.west) -- (localize);
        \path [draw, ->, shorten >= 1.5pt, shorten <= 1.5pt] (delta_edb) to [bend right=35] (localize.south);

        \path [draw, ->, shorten >= 1.5pt, shorten <= 1.5pt, dashed] (localize) to [bend left=15] (repair.west);
        \path [draw, ->, shorten >= 1.5pt, shorten <= 1.5pt, dashed] (repair.east) to [bend left=15] (person_head.north);
        \path [draw, ->, shorten >= 1.5pt, shorten <= 1.5pt, dashed] (person_feet.south) to [bend left=15] (feedback.east);
        \path [draw, ->, shorten >= 1.5pt, shorten <= 1.5pt, dashed] (feedback.west) to [bend left=15] (localize);

        \draw [|-|] (-1, 2.5) -- node[above]{Pre-processing Incremental Datalog evaluation} (11.7, 2.5);
        \draw [|-|] (12.3, 2.5) -- node[above]{\textbf{Fault Localization/Repair}} (19, 2.5);
    \end{tikzpicture}
    \caption{Fault Localization and Repair System\label{fig:system}}
\end{figure}

An overview of our approach is shown in Figure~\ref{fig:system}. The first
portion of the system is the incremental Datalog evaluation. Here, the
incremental evaluation takes an EDB and an incremental update containing tuples
inserted or deleted from the EDB, denoted $\Delta$EDB. The result of the
incremental evaluation is the output IDB, along with the set of IDB insertions
and deletions from the incremental update, denoted $\Delta$IDB. The evaluation
also enables incremental provenance, producing a proof tree for a given query
tuple.

The second portion of the system is the fault localization/rollback repair. This
process takes a set of faults provided by the user, which is a subset of
$\Delta$IDB where each tuple is either unwanted and inserted in $\Delta$IDB or
is desirable but deleted in $\Delta$IDB. Then, the fault localization and
rollback repair algorithms use the full $\Delta$IDB and $\Delta$EDB, along with
incremental provenance, to produce a localization or rollback suggestion.

The main fault localization and rollback algorithms work in tandem to provide
localizations or rollback suggestions to the user. The key idea of these
algorithms is to compute proof trees for fault tuples using the provenance
utility provided by the incremental Datalog engine. These proof trees directly
provide localization for the faults. For fault rollback, the algorithms create
an Integer Linear Programming (ILP) instance that encodes the proof trees, with
the goal of \emph{disabling} all proof trees to prevent the fault tuples from
appearing.

The result is a localization or rollback suggestion, which is a subset of
$\Delta$EDB. For localization, the subset $S \subseteq \Delta \EDB$ is such that
if we were to apply $S$ to $\EDB$ as the diff, the set of faults would be
reproduced. For a rollback suggestion, the subset $S \subseteq \Delta \EDB$ is
such that if we were to remove $S$ from $\Delta$EDB, then the resulting diff
would \emph{not} produce the faults.

%% file: src/inc-prov.tex
\section{Incremental Provenance}
\label{sec:inc-prov}

\subsection{Background} \label{sec:inc-prov-background}

\paragraph{Provenance.} \label{para:provenance}
Provenance~\cite{caballero2017survey,zhao2020debugging,raghothaman2019provenance}
provides machinery to explain the existence of a tuple. For example, the tuple
\texttt{vpt(userSession, L3)} could be explained in our running example by the
following proof tree:
\begin{prooftree}
    \AxiomC{$\mathtt{assign}(\mathtt{userSession}, \mathtt{ins})$}
    \AxiomC{$\mathtt{new}(\mathtt{ins}, \mathtt{L3})$}

    \RightLabel{$r_1$}
    \UnaryInfC{$\mathtt{vpt}(\mathtt{ins}, \mathtt{L3})$}

    \RightLabel{$r_2$}
    \BinaryInfC{$\mathtt{vpt}(\mathtt{userSession}, \mathtt{L3})$}
    \vspace{1em}
\end{prooftree}

This proof tree explains the derivation of \texttt{vpt(userSession, L3)} through
input and intermediate tuples and which Datalog rules were involved in the
derivation. Therefore, these proof trees provide a link between faulty tuples
and the structure of the program itself. This forms the basis for our incremental Datalog
debugging approach.

A two-phase approach for computing provenance was introduced in~\cite{zhao2020debugging}.
The key idea is to instrument the Datalog program to compute some provenance
information alongside the actual tuples, and then to use this information to
produce proof trees.

The two-phase process consists of:
\begin{enumerate}
    \item Instrumented Datalog evaluation: annotating each tuple with
    \emph{provenance annotations} while computing the IDB. In particular, for each
    tuple $t$, the system stores the height of the minimal height proof tree for
    $t$. For an EDB tuple, the height is $0$, meanwhile the height of an IDB tuple
    is computed by $h(t) = \max \{ h \brackets{t_1}, \ldots, h \brackets{t_k}
    \} + 1$ if $t$ is derived by a rule instantiation $t \dlimpl t_1, \ldots, t_k$.

    \item Provenance query answering: using the annotated IDB to answer provenance
    queries of the form of ``\texttt{explain vpt(userSession, L3)}''. Internally,
    the system generates and solves constraints using the provenance annotations to
    construct the proof trees one level at a time.
\end{enumerate}

For the running example, the tuple \texttt{vpt(userSession, L3)} gets a height
annotation of $2$ during the instrumented evaluation phase. Then, during the query
answering, the system generates a constraint saying ``find the tuples matching the body of a
rule for \texttt{vpt} with height less than $2$''.


\paragraph{Incremental Evaluation.} \label{para:inceval}
Incremental Datalog evaluation provides an efficient method of \emph{updating} the results
of a computation given some small changes to the input (where some tuples are inserted and/or deleted). Previous incremental evaluation
approaches, such as Delete-Rederive (DRed)~\cite{gupta1993maintaining}, had shortcomings concerning
over-deletion and the re-derivation step required to resolve this. However, modern
incremental evaluation approaches typically use a counting-based approach. For these
strategies, each IDB tuple is associated with a count representing the number of
different derivations for that tuple. When an EDB tuple $t$ is inserted or deleted, tuples
depending on $t$ have their counts incremented or decremented respectively. If a
tuple's count reaches 0, then that tuple is removed from the IDB.

To handle recursion, techniques such as Differential Dataflow~\cite{mcsherry2013differential}
propose storing counters for each tuple \emph{per iteration} of the recursion. Note that if rules are alternately increasing and decreasing, then they would be operating in different iterations.
Therefore, each tuple is associated with an iteration number and a count. This
increases the space overhead for keeping the auxiliary information but allows for
precise recording of the incremental updates in a recursive setting. Variations of this idea, such as
Elastic Incremental Evaluation~\cite{zhao2021towards}, propose a trade-off of
lower space overhead at the cost of requiring some recomputation to maintain
precision of the derivation counts.


In our running example, the
insertion of two lines in the source program result in the insertion of two EDB tuples
for the pointer analysis: \texttt{assign(upgradedSession, userSession)} and
\texttt{load(userSession, admin, session)}.

Using a counting approach~\cite{zhao2021towards,mcsherry2013differential}, we can apply incremental counting
versions of the Datalog rules to compute an updated IDB. In this instance, we
can apply rule $r_2$ as
\begin{Verbatim}[commandchars=\\\{\}]
    vpt(upgradedSession,L3)\textsuperscript{+1} :-
        assign(upgradedSession,userSession)\textsuperscript{+1},
        vpt(userSession,L3).
\end{Verbatim}

The superscripts denote the changes in count as a result of the newly inserted
EDB tuple \texttt{assign(upgradedSession,userSession)}. These updated counts
are then propagated in subsequent rules, until the IDB has reached fixpoint. Further details are described in \cite{zhao2021towards}.


\subsection{Combining Provenance and Incremental Evaluation}

\begin{figure*}
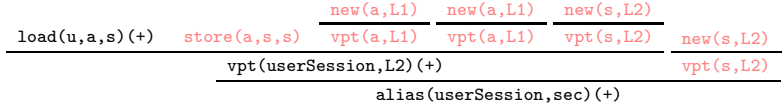

    \begin{prooftree}
        \def\defaultHypSeparation{\hskip .1cm}
        \AxiomC{\scriptsize{\texttt{load(u,a,s)(+)}}}
        \AxiomC{\textcolor{red!50}{\scriptsize{\texttt{store(a,s,s)}}}}
        \AxiomC{\textcolor{red!50}{\scriptsize{\texttt{new(a,L1)}}}}
        \UnaryInfC{\textcolor{red!50}{\scriptsize{\texttt{vpt(a,L1)}}}}
        \AxiomC{\textcolor{red!50}{\scriptsize{\texttt{new(a,L1)}}}}
        \UnaryInfC{\textcolor{red!50}{\scriptsize{\texttt{vpt(a,L1)}}}}
        \AxiomC{\textcolor{red!50}{\scriptsize{\texttt{new(s,L2)}}}}
        \UnaryInfC{\textcolor{red!50}{\scriptsize{\texttt{vpt(s,L2)}}}}
        \QuinaryInfC{\scriptsize{\texttt{vpt(userSession,L2)(+)}}}
        \AxiomC{\textcolor{red!50}{\scriptsize{\texttt{new(s,L2)}}}}
        \UnaryInfC{\textcolor{red!50}{\scriptsize{\texttt{vpt(s,L2)}}}}
        \BinaryInfC{\scriptsize{\texttt{alias(userSession,sec)(+)}}}
    \end{prooftree}
    \caption{The proof tree for \texttt{alias(userSession,sec)}. \texttt{(+)}
    denotes tuples that are inserted as a result of the incremental update, red
    denotes tuples that were not affected by the incremental update.}
    \label{fig:inc-proof-tree}
\end{figure*}

For fault localization and rollback, a novel provenance strategy is required
that builds on incremental evaluation. \emph{Incremental provenance} restricts
the computations of the proof tree to the portions affected by the incremental
update only. For example, Figure~\ref{fig:inc-proof-tree} shows an incremental
proof tree for the inserted tuple \texttt{alias(userSession,sec)}. The tuples
labeled with \texttt{(+)} are inserted by an incremental update. Incremental
provenance would only compute provenance information for these newly inserted
tuples and would not explore the tuples in red already established in a previous
epoch.

To formalize incremental provenance, we define $\prov$ as follows. Given an
incremental update $\Delta E$, $\prov(P, E, t, \Delta E)$ should consist of tuples
that were updated due to the incremental update.
\begin{definition}
    Let $P$ be a Datalog program, $E$ be an EDB,
    $\Delta E$ be an incremental update, and $t$ be a tuple contained in both
    $E$ and $\Delta E$. 
    Then, $\prov(P, E, t, \Delta E)$ is the set of tuples that appear in the proof
    tree for a tuple $t$, that are also inserted as a result of $\Delta E$. In
    the remainder of the paper, we omit $P$ and $E$ if they are unambiguous.
\end{definition}



To compute provenance information efficiently in an incremental
evaluation setting, we introduce a novel method combining the provenance
annotations of \cite{zhao2020debugging} with the incremental annotations of
\cite{zhao2021towards}. Recall, from Section~\ref{sec:inc-prov-background}, that provenance annotations include the height of
the minimal height proof tree for a tuple, computed by taking the maximum height
of all tuples in its derivation. Also recall that incremental annotations
include the iteration number in which a tuple is derived.

To combine these two, we can observe a correspondence between the iteration number and
provenance annotations. A tuple is produced in some iteration if at least one of
the body tuples was produced in the previous iteration. Therefore, the iteration
number $I$ for a tuple produced in a fixpoint is equivalent to
\begin{align*}
    I(t) = \max \{ I(t_1), \ldots, I(t_k) \} + 1
\end{align*}
if $t$ is computed by rule instantiation $t \dlimpl t_1, \ldots, t_k$. This
definition of iteration number corresponds closely to the height annotation in
provenance. Therefore, the iteration number is suitable for constructing proof
trees similar to provenance annotations.

For fault localization and rollback, it is also important that the Datalog
engine produces only provenance information that is \emph{relevant} for faults
that appear after an incremental update. Therefore, the provenance information
produced by the Datalog engine should be restricted to tuples inserted or
deleted by the incremental update. Thus, we adapt the user-driven proof tree
exploration process in~\cite{zhao2020debugging} to use an automated procedure
that enumerates exactly the portions of the proof tree that have been affected
by the incremental update. 

As a result, our approach for incremental provenance produces proof trees
containing only tuples inserted or deleted due to an update. For fault
localization and rollback, this property is crucial for minimizing the search
space when computing localizations and rollback suggestions.

%% file: src/fault-localization.tex
\section{Fault Localization and Rollback Repair}
\label{sec:fault-localization}

This section describes our approach and algorithms for both the fault
localization and rollback problems. We begin by formalizing the problem and then
presenting basic versions of both problems. Finally, we extend the algorithms to
handle missing faults and negation.

\subsection{Preliminaries}
We first define a fault to formalize the fault localization and rollback
problems. For a Datalog program, a fault may manifest as either (1) an
undesirable tuple that appears or (2) a desirable tuple that disappears. In
other words, a fault is a tuple that does not match the \emph{intended
output} of a program.

\begin{definition}[Intended Output]
    The \emph{intended output} of a Datalog program $P$ is a pair of sets
    $(I_+, I_-)$ where $I_+$ and $I_-$ are desirable and undesirable tuple sets,
    respectively. An input set $E$ is \emph{correct} w.r.t $P$ and $(I_+, I_-)$
    if $I_+ \subseteq P(E)$ and $I_- \cap P(E) = \emptyset$.
\end{definition}

Given an intended output for a program, a \emph{fault} can be defined as
follows:

\begin{definition}[Fault]
    Let $P$ be a Datalog program, with input set $E$ and intended output
    $(I_+, I_-)$. Assume that $E$ is incorrect w.r.t. $P$ with $(I_+, I_-)$.
    Then, a \emph{fault} of $E$ is a tuple $t$ such that either $t$ is desirable
    but missing, i.e., $t \in I_+ \setminus P(E)$ or $t$ is undesirable but
    produced, i.e., $t \in P(E) \cap I_-$.
\end{definition}

We can formalize the situation where an incremental update for a Datalog program
introduces a fault. Let $P$ be a Datalog program with intended output
$I_{\checkmark} = (I_+, I_-)$ and let $E_1$ be an input EDB.  Then, let $\Delta
E_{1 \rightarrow 2}$ be an incremental update (or \emph{diff}), such that the application
operator $E_1 \uplus \Delta E_{1 \rightarrow 2}$ results in another input EDB,
$E_2$. Then, assume that $E_1$ is
correct w.r.t $I_{\checkmark}$, but $E_2$ is incorrect.

\paragraph{Fault Localization.} The fault localization problem allows the
user to pinpoint the sources of faults. This is achieved by providing a
minimal subset of the incremental update that can still reproduce the fault.

\begin{definition}[Fault Localization]
    A \emph{fault localization} is a subset $\delta E \subseteq \Delta
    E_{1 \rightarrow 2}$ such that $P(E_1 \uplus \delta E)$ exhibits \emph{all}
    faults of $E_2$.
\end{definition}

\paragraph{Rollback Suggestion.} A rollback suggestion provides a subset of the
diff, such that its removal from the diff would fix all faults. 
\begin{definition}[Rollback Suggestion]
    A \emph{rollback suggestion} is a subset $\delta E_{\times}
    \subseteq \Delta E_{1 \rightarrow 2}$ such that $P(E_1 \uplus (\Delta E_{1
    \rightarrow 2} \setminus \delta E_{\times}))$ does not produce any faults of
    $E_2$.
\end{definition}


\subsection{Fault Localization}
In the context of incremental Datalog, the \emph{fault localization problem}
provides a small subset of the incremental changes that allow the fault to be
reproduced.
On its own, fault localization forms an important part of the
reproduction step of any fault investigation. Moreover, it is also fundamental
for rollback repair of missing tuples or negations (see
Section~\ref{sec:algo-extensions}).

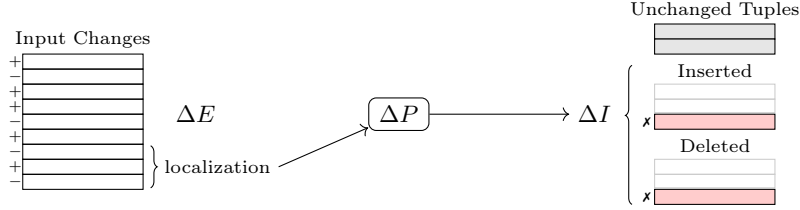
\begin{figure}[h]
    \centering
    \begin{tikzpicture}
        \node[shape=rectangle, draw=black, minimum width=0.8cm, rounded corners=0.1cm] (P1) at (3,1) {$\Delta P$};



        \node[] (delta) at (0.3,1) {$\Delta E$};


        \node[anchor=center] at (-1.2,2) {\scriptsize Input Changes};
        \draw[] (-2,1.8) rectangle (-0.4,1.6);
        \draw[] (-2,1.6) rectangle (-0.4,1.4);
        \draw[] (-2,1.4) rectangle (-0.4,1.2);
        \draw[] (-2,1.2) rectangle (-0.4,1.0);
        \draw[] (-2,1.0) rectangle (-0.4,0.8);
        \draw[] (-2,0.8) rectangle (-0.4,0.6);
        \draw[] (-2,0.6) rectangle (-0.4,0.4);
        \draw[] (-2,0.4) rectangle (-0.4,0.2);
        \draw[] (-2,0.2) rectangle (-0.4,0.0);

        \node[anchor=center] at (-2.1,1.7) {\tiny $+$};
        \node[anchor=center] at (-2.1,1.5) {\tiny $-$};
        \node[anchor=center] at (-2.1,1.3) {\tiny $+$};
        \node[anchor=center] at (-2.1,1.1) {\tiny $+$};
        \node[anchor=center] at (-2.1,0.9) {\tiny $-$};
        \node[anchor=center] at (-2.1,0.7) {\tiny $+$};
        \node[anchor=center] at (-2.1,0.5) {\tiny $-$};
        \node[anchor=center] at (-2.1,0.3) {\tiny $+$};
        \node[anchor=center] at (-2.1,0.1) {\tiny $-$};

        \draw [decorate,decoration={brace}] (-0.3,0.58) -- (-0.3,0.02)
        node[pos=0.5,right=2pt,black,name=l] {\scriptsize localization};

        \path[->] (l.east) edge (P1);

        \node[] (I) at (5.6,1) {$\Delta I$};
        \path[->] (P1) edge (I);

        \node[anchor=center] at (7.2,2.38) {\scriptsize Unchanged Tuples};
        \draw[fill=black!10] (6.4,2.2) rectangle (8,2.0);
        \draw[fill=black!10] (6.4,2.0) rectangle (8,1.8);
        \node[anchor=center] at (7.2,1.58) {\scriptsize Inserted};
        \draw[draw=black!20] (6.4,1.4) rectangle (8,1.2);
        \draw[draw=black!20] (6.4,1.2) rectangle (8,1.02);
        \draw[fill=red!20] (6.4,1.0) rectangle (8,0.8);
        \node[anchor=center] at (7.2,0.58) {\scriptsize Deleted};
        \draw[draw=black!20] (6.4,0.4) rectangle (8,0.2);
        \draw[draw=black!20] (6.4,0.2) rectangle (8,0.02);
        \draw[fill=red!20] (6.4,0.0) rectangle (8,-0.2);

        \draw[decorate,decoration={brace,aspect=0.64}] (6.1,-0.2) -- (6.1,1.65);




        \node[anchor=center] at (6.3,0.9) {\tiny \ding{55}};
        \node[anchor=center] at (6.3,-0.1) {\tiny \ding{55}};

    \end{tikzpicture}
    \caption{A fault localization is a subset of input changes such that the
    faults are still reproduced}
    \label{fig:incprov:faultlocalization}
\end{figure}

Consider the example in Figure~\ref{fig:incprov:faultlocalization}.
This diagram illustrates that a fault localization is a subset of the input
changes $L \subseteq \Delta E$ such that when $L$ is used as the input changes
in the incremental evaluation, the resulting update still produces the faults.

We begin by considering a basic version of the fault localization problem. In
this basic version, we have a positive Datalog program (i.e., with no negation),
and we localize a set of faults that are undesirable but appear (i.e., $P(E)
\cap I_-$). The main idea of the fault localization algorithm is to compute a
proof tree for each fault tuple. The tuples forming these proof trees are
sufficient to localize the faults since these tuples allow the proof trees to be
valid and, thus, the fault tuples to be reproduced.

\vspace{-1em}
\begin{algorithm}[H]
    \caption{Localize-Faults($P$, $E_2$, $\Delta E_{1 \rightarrow 2}$, $F$): Given a diff
    $\Delta E_{1 \rightarrow 2}$ and a set of fault tuples $F$, returns $\delta E \subseteq
    \Delta E_{1 \rightarrow 2}$ such that $E_1 \uplus \delta E$ produces all $t \in F$}
    \label{alg:localize-fault}

    \begin{algorithmic}[1]
        \For{tuple $t \in F$}
            \State{Let $\prov(P, E2, t, \Delta E_{1 \rightarrow 2})$ be an incremental proof tree of $t$ containing tuples that were inserted due to $\Delta E_{1 \rightarrow 2}$}
        \EndFor
        \State \Return{$\cup_{t \in F} (\prov(t, \Delta E_{1 \rightarrow 2}) \cap \Delta E_{1 \rightarrow 2})$}
    \end{algorithmic}
\end{algorithm}
\vspace{-1em}

The basic fault localization is presented in Alg.~\ref{alg:localize-fault}. For
each fault tuple $t \in F$, the algorithm computes one incremental proof tree
$\prov(t, \Delta E_{1 \rightarrow 2})$. These proof trees contain the set of
tuples that were inserted due to the incremental update $\Delta E_{1 \rightarrow
2}$ and cause the existence of each fault tuple $t$. Therefore, by returning
the union $\cup_{t \in F} (\prov(t, \Delta E_{1 \rightarrow 2}) \cap \Delta E_{1
\rightarrow 2})$, the algorithm produces a subset of $\Delta E_{1 \rightarrow
2}$ that reproduces the faults.

\subsection{Rollback Repair}
\begin{figure}[h]
    \centering
    \begin{tikzpicture}
        \node[shape=rectangle, draw=black, minimum width=0.8cm, rounded corners=0.1cm] (P1) at (3,1) {$\Delta P$};





        \node[anchor=center] at (-1.2,2) {\scriptsize Input Changes};
        \draw[] (-2,1.8) rectangle (-0.4,1.6);
        \draw[] (-2,1.6) rectangle (-0.4,1.4);
        \draw[] (-2,1.4) rectangle (-0.4,1.2);
        \draw[] (-2,1.2) rectangle (-0.4,1.0);
        \draw[] (-2,1.0) rectangle (-0.4,0.8);
        \draw[] (-2,0.8) rectangle (-0.4,0.6);
        \draw[] (-2,0.6) rectangle (-0.4,0.4);
        \draw[] (-2,0.4) rectangle (-0.4,0.2);
        \draw[] (-2,0.2) rectangle (-0.4,0.0);

        \node[anchor=center] at (-2.1,1.7) {\tiny $+$};
        \node[anchor=center] at (-2.1,1.5) {\tiny $-$};
        \node[anchor=center] at (-2.1,1.3) {\tiny $+$};
        \node[anchor=center] at (-2.1,1.1) {\tiny $+$};
        \node[anchor=center] at (-2.1,0.9) {\tiny $-$};
        \node[anchor=center] at (-2.1,0.7) {\tiny $+$};
        \node[anchor=center] at (-2.1,0.5) {\tiny $-$};
        \node[anchor=center] at (-2.1,0.3) {\tiny $+$};
        \node[anchor=center] at (-2.1,0.1) {\tiny $-$};

        \draw [decorate,decoration={brace}] (-0.3,0.78) -- (-0.3,0.02)
        node[pos=0.5,right=2pt,black] {\scriptsize rollback repair};

        \draw [decorate,decoration={brace}] (-0.3,1.78) -- (-0.3,0.82)
        node[pos=0.5,right=0.1pt,black,name=l] {};

        \path[->] (l.east) edge (P1);

        \node[] (I) at (5.6,1) {$\Delta I$};
        \path[->] (P1) edge (I);

        \node[anchor=center] at (7.2,2.38) {\scriptsize Unchanged Tuples};
        \draw[fill=black!10] (6.4,2.2) rectangle (8,2.0);
        \draw[fill=black!10] (6.4,2.0) rectangle (8,1.8);
        \node[anchor=center] at (7.2,1.58) {\scriptsize Inserted};
        \draw[fill=green!20] (6.4,1.4) rectangle (8,1.2);
        \draw[fill=green!20] (6.4,1.2) rectangle (8,1.02);
        \draw[draw=black!20] (6.4,1.0) rectangle (8,0.8);
        \node[anchor=center] at (7.2,0.58) {\scriptsize Deleted};
        \draw[fill=green!20] (6.4,0.4) rectangle (8,0.2);
        \draw[fill=green!20] (6.4,0.2) rectangle (8,0.02);
        \draw[draw=black!20] (6.4,0.0) rectangle (8,-0.2);

        \draw[decorate,decoration={brace,aspect=0.64}] (6.1,-0.2) -- (6.1,1.65);




        \node[anchor=center] at (6.3,0.9) {\tiny \ding{55}};
        \node[anchor=center] at (6.3,-0.1) {\tiny \ding{55}};

    \end{tikzpicture}
    \caption{An input debugging suggestion is a subset of input changes such
    that the remainder of the input changes no longer produce the faults}
    \label{fig:incprov:debugsuggestion}
\end{figure}
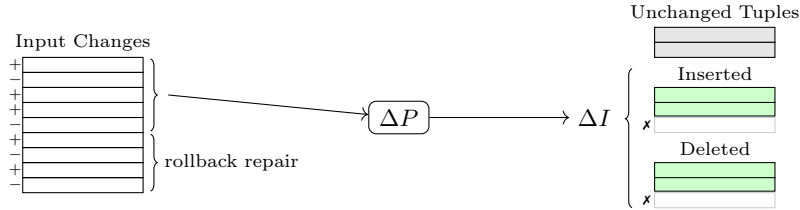

A rollback repair is a subset of the input changes such that the
remaining changes `fix' the faults. Consider
Figure~\ref{fig:incprov:debugsuggestion}, which shows that a rollback repair is
a small subset of the input changes, such that the remainder of the changes no longer produce the faults when used
as an incremental update.

The rollback repair algorithm produces a rollback suggestion. As with fault
localization, our presentation begins with a basic version of the rollback problem, where we
have only a positive Datalog program and wish to roll back a set of unwanted
fault tuples. The basic rollback repair algorithm involves computing \emph{all}
non-cyclic proof trees for each fault tuple and `disabling' each of those proof
trees, as shown in Alg.~\ref{alg:repair}. If all proof trees are invalid, the
fault tuple will no longer be computed by the resulting EDB.

\begin{algorithm}
    \caption{Rollback-Repair($P$, $E_2$, $\Delta E_{1 \rightarrow 2}$, $F$): Given a
    diff $\Delta E_{1 \rightarrow 2}$ and a set of fault tuples $F$, return a subset $\delta E
    \subseteq \Delta E_{1 \rightarrow 2}$ such that $E_1 \uplus (\Delta E_{1 \rightarrow 2} \setminus \delta E)$
    does not produce $t_r$}
    \label{alg:repair-fault}

    \begin{algorithmic}[1]
        \State{Let $\allprov(t, \Delta E_{1 \rightarrow 2}) = \{T_1, \ldots, T_n\}$ be the total incremental provenance
        for a tuple $t$ w.r.t $P$ and $E_2$, where each $T_i$ is a non-cyclic proof tree containing tuples inserted due
        to $\Delta E_{1 \rightarrow 2}$.}

        \Statex{\textit{Construct an integer linear program instance as follows:}}

        \State{Create a $0/1$ integer variable $x_{t_k}$ for each tuple $t_k$ that occurs
        in the proof trees in $\allprov(t, \Delta E_{1 \rightarrow 2})$ for each fault tuple $t \in F$}

        \For{each tuple $t_f \in F$}
            \For{each proof tree $T_i \in \allprov(t_f, \Delta E_{1 \rightarrow 2})$}
                \For{each line $t_h \leftarrow t_1 \land \ldots \land t_k$ in $T_i$}
                    \State{Add a constraint $x_{t_1} + \ldots + x_{t_k} - x_{t_h} \leq k - 1$}
                \EndFor
            \EndFor
            \State{Add a constraint $x_{t_f} = 0$}
        \EndFor

        \State{Add the objective function $\text{maximize} \sum_{{t_e} \in EDB} x_{t_e}$}

        \State{Solve the ILP}
        \State{Return $\{ t \in \Delta E_{1 \rightarrow 2} \mid x_t = 0\}$}
    \end{algorithmic}
    \label{alg:repair}
\end{algorithm}

Alg.~\ref{alg:repair-fault} computes a minimum subset of the diff $\Delta E_{1
\rightarrow 2}$, which would prevent the production of each $t \in F$ when
excluded from the diff. The key idea is to use integer linear programming
(ILP)~\cite{ILP} as a vehicle to disable $\Delta$EDB tuples so that the fault tuples
vanish in the resulting IDB. We phrase the proof trees as a pseudo-Boolean
formula~\cite{Hooker} whose propositions represent the $\Delta$EDB and IDB tuples in the proof trees. In
the ILP, the faulty tuples are constrained to be false, and the $\Delta$EDB tuples
assuming the true value are to be maximized, i.e., we wish to eliminate the
least number of tuples in $\Delta$EDB for repair. The ILP created in
Alg.~\ref{alg:repair-fault} introduces a variable for each tuple (either IDB or
$\Delta$EDB) that appears in \emph{all} incremental proof trees for the fault tuples.
For the ILP model, we have three types of constraints: 
\begin{enumerate}
    \item to encode each one-step proof tree,
    \item to enforce that fault tuples are false, and
    \item to ensure that variables are in the $0$-$1$ domain.
\end{enumerate}
The constraints encoding
proof trees correspond to each one-step derivation which can be expressed as a
Boolean constraint, where $t_1, \ldots, t_k$ and $t_h$ are Boolean variables:
\begin{align*}
    \frac{t_1\ \ \ldots\ \ t_k}{t_h}\ \equiv\ {t_1} \land \ldots \land {t_k} \implies {t_h}
\end{align*}
Using
propositional logic rules, this is equivalent to $\overline{t_1} \lor \ldots
\lor \overline{t_k} \lor {t_h}$. This formula is then transformed into a pseudo
(linear) Boolean formula where $\varphi$ maps a Boolean function to the $0-1$
domain, and $x_t$ corresponds to the $0$-$1$ variable of proposition $t$ in the
ILP.
\begin{align*}
    \varphi\left( \overline{t_1} \lor \ldots \lor \overline{t_k} \lor {t_h} \right) &\equiv 
    (1-x_{t_1}) + \ldots + (1-x_{t_k}) + t_h > 0  \\
    &\equiv x_{t_1} + \ldots + x_{t_k} - x_{t_h} \leq k - 1
\end{align*}

The constraints assuming false values for fault tuples $t_f \in F$ are simple
equalities, i.e., $x_{t_f} = 0$. The objective function for the ILP is to
maximize the number of inserted tuples that are kept, which is equivalent to
minimizing the number of tuples in $\Delta E_{1 \rightarrow 2}$ that are
disabled by the repair. In ILP form, this is expressed as maximizing $\sum_{t
\in \Delta E_{1 \rightarrow 2}} t$.

\begin{equation*}
\begin{array}{lll}
    \mbox{max.}  & \sum_{t \in \Delta E_{1 \rightarrow 2}} x_{t} \\
    \mbox{s.t.} & x_{t_1} + \ldots x_{t_k} - x_{t_h} \leq k -1 & \left(\forall\ \frac{t_1\ \ \ldots\ \ t_k}{t_h} \in T_i \right) \\
                & x_{t_f} = 0 & \left( \forall t_f \in F \right)\\
                & x_t \in \{0, 1\} & \left( \forall \text{tuples } t \right)
\end{array}
\end{equation*}

The solution of the ILP permits us to determine the EDB tuples for repair, i.e., $\{t \in \Delta E_{1 \rightarrow 2} \mid x_t = 0\}$ 
This is a minimal set of inserted tuples that must be removed from $\Delta E_{1
\rightarrow 2}$ so that the fault tuples disappear.

This ILP formulation encodes the problem of disabling all proof trees for all
fault tuples while maximizing the number of inserted tuples kept in the result.
If there are multiple fault tuples, the algorithm computes proof trees for each
fault tuple and combines all proof trees in the ILP encoding. The result is a
set of tuples that is minimal but sufficient to prevent the fault tuples from
being produced.

\subsection{Extensions}
\label{sec:algo-extensions}

\paragraph{Missing Tuples.} The basic versions of the fault localization and
rollback repair problem only handle a tuple which is undesirable but appears.
The opposite kind of fault, i.e., a tuple which is desirable but missing, can be
localized or repaired by considering a \emph{dual} version of the problem. For
example, consider a tuple $t$ that disappears after applying a diff $\Delta
E_{1 \rightarrow 2}$, and appears in the update in the opposite direction,
$\Delta E_{2 \rightarrow 1}$. Then, the dual problem of localizing the
disappearance of $t$ is to \emph{roll back} the appearance of $t$ after applying
the opposite diff, $\Delta E_{2 \rightarrow 1}$.

To localize a disappearing tuple $t$, we want to provide a small subset $\delta
E$ of $\Delta E_{1 \rightarrow 2}$ such that $t$ is still not computable after
applying $\delta E$ to $E_1$. To achieve this, \emph{all} ways to derive $t$
must be invalid after applying $\delta E$. Considering the dual problem, rolling
back the appearance of $t$ in $\Delta E_{2 \rightarrow 1}$ results in a subset
$\delta E$ such that $E_2 \uplus (\Delta E_{2 \rightarrow 1} \setminus \delta
E)$ does not produce $t$. Since $E_1 = E_2 \uplus \Delta E_{2 \rightarrow 1}$,
if we were to apply the reverse of $\delta E$ (i.e., insertions become deletions
and vice versa), we would arrive at the same EDB set as $E_2 \uplus (\Delta E_{2
\rightarrow 1} \setminus \delta E)$. Therefore, the reverse of $\delta E$ is the
desired minimal subset that localizes the disappearance of $t$.

Similarly, to roll back a disappearing tuple $t$, we apply the dual problem of
\emph{localizing} the appearance of $t$ after applying the opposite diff $\Delta
E_{2 \rightarrow 1}$. Here, to roll back a disappearing tuple, we introduce
\emph{one} way of deriving $t$. Therefore, localizing the appearance of $t$ in
the opposite diff provides one derivation for $t$ and thus is the desired
solution. In summary, to localize or rollback a tuple $t$ that is missing after
applying $\Delta E_{1 \rightarrow 2}$, we compute a solution for the dual
problem. The dual problem for localization is to roll back the appearance of $t$
after applying $\Delta E_{2 \rightarrow 1}$, and similarly, the dual problem for
rollback is localization. We note the fault localization on its own is an important part of the investigation process, in case algorithm will compute a fault localization for the diff in the reverse direction.

\paragraph{Stratified Negation.} Stratified negation is a common extension for
Datalog. With stratified negation, atoms in the body of a Datalog rule may
appear negated, e.g., 
$$
    R_h(X_h) \dlimpl R_1(X_1), \ldots, !R_k(X_k), \ldots, R_n(X_n).
$$
The negated atoms are denoted with $!$, and any variables contained in negated
atoms must also exist in some positive atom in the body of the rule (a property
called \emph{groundedness} or \emph{safety}). Semantically, negations are satisfied if the
instantiated tuple \emph{does not} exist in the corresponding relation. The
`stratified' in `stratified negation' refers to the property that no cyclic
negations can exist. For example, the rule $A(x) \dlimpl B(x, y), !A(y)$ causes
a dependency cycle where $A$ depends on the negation of $A$ and thus is not
allowed under stratified negation.

Consider the problem of localizing the appearance of an unwanted tuple $t$. If
the Datalog program contains stratified negation, then the appearance of $t$ can
be caused by two possible situations. Either (1) there is a positive tuple in
the proof tree of $t$ that appears, or (2) there is a negated tuple in the proof
tree of $t$ that disappears. The first case is the standard case, but in the
second case, if a negated tuple disappears, then its disappearance can be
localized or rolled back by computing the dual problem as in the missing tuple
strategy presented above. We may encounter further negated tuples in executing
the dual version of the problem. For example, consider the set of Datalog rules
$A(x) \dlimpl B(x), !C(x)$ and  $C(x) \dlimpl D(x), !E(x)$. If we wish to
localize an appearing  (unwanted) tuple \texttt{A(x)}, we may encounter a
disappearing tuple \texttt{C(x)}. Then, executing the dual problem, we may
encounter an appearing tuple \texttt{E(x)}. We can generally continue flipping
between the dual problems to solve the localization or repair problem. This
process is guaranteed to terminate due to the stratification of negations. Each
time the algorithm encounters a negated tuple, it must appear in an earlier
stratum than the previous negation. Therefore, eventually, the negations will
reach the input EDB, and the process terminates.

\paragraph{Changes in Datalog Rules.} The algorithms are presented above in the
context of localizing or debugging a change to the input tuples. However, with
a simple transformation, the same algorithms can also be applied to changes in Datalog
rules. For each Datalog rule, introduce a predicate \texttt{Rule(i)}, where \texttt{i}
is a unique number per rule. Then, the unary relation \texttt{Rule} can be
considered as EDB, and thus the set of rules can be changed by providing
a diff containing insertions or deletions into the \texttt{Rule} relation.
For example, a transformed set of rules may be:
\begin{verbatim}
    P(x, y) :- E(x, y), Rule(1).
    P(x, z) :- E(x, y), P(y, z), Rule(2).
\end{verbatim}
Then, by including or excluding 1 or 2 in the EDB relation \texttt{Rule}, the
underlying Datalog rules can be `switched on or off,' and a change to the
Datalog program can be expressed as a diff in the \texttt{Rule} relation.

\subsection{Full Algorithm}

\begin{algorithm}
    \caption{Full-Rollback-Repair($P$, $E_1$, $\Delta E_{1 \rightarrow 2}$, $(I_+,
    I_-)$): Given a diff $\Delta E_{1 \rightarrow 2}$ and an intended output
    $(I_+, I_,)$, compute a subset $\delta E \subseteq \Delta E_{1 \rightarrow 2}$
    such that $\Delta E_{1 \rightarrow 2} \setminus \delta E$ satisfies the intended
    output}
    \label{alg:full-fault-repair}

    \begin{algorithmic}[1]
        \State{Let $E_2$ be the EDB after applying the diff: $E_1 \uplus \Delta
        E_{1 \rightarrow 2}$}\label{line:initialize-edb}
        \State{Let $F^+$ be appearing unwanted faults: $\{I_- \cap P(E_2) \}$}\label{line:initialize-fault-1}
        \State{Let $F^-$ be missing desirable faults: $\{I_+ \setminus P(E_2) \}$}\label{line:initialize-fault-2}

        \State{Let $L$ be the set of repair tuples, initialized to
        $\emptyset$} \label{line:initialize-localization}

        \While{both $F^+$ and $F^-$ are non-empty} \label{line:localize-loop-start}
            \State{Add Rollback-Repair($P$, $E_2$, $\Delta E_{1 \rightarrow 2}$, $F^+$) to $L$} \label{line:repair-t+}

            \For{negated tuples $!t \in L$} \label{line:add-!t-to-F-}
                \State{Add $t$ to $F^-$}
            \EndFor

            \State{Clear $F^+$}

            \State{Add Localize-Faults($P$, $E_1$, $\Delta E_{2 \rightarrow 1}$, $F^-$) to $L$} \label{line:localize-t-}

            \For{negated tuples $!t \in L$}
                \State{Add $t$ to $F^+$}
            \EndFor

            \State{Clear $F^-$} \label{line:clear-f-}
        \EndWhile

        \State{\Return $L$}
    \end{algorithmic}
\end{algorithm}

The full rollback repair algorithm presented in Alg.~\ref{alg:full-fault-repair}
incorporates the basic version of the problem and all of the extensions
presented above. The result of the algorithm is a rollback suggestion, which
fixes all faults. Alg.~\ref{alg:full-fault-repair} begins by initializing the
EDB after applying the diff (line~\ref{line:initialize-edb}) and separate sets
of unwanted faults (lines~\ref{line:initialize-fault-1}) and missing faults
(\ref{line:initialize-fault-2}). The set of candidate tuples forming the repair
is initialized to be empty (line~\ref{line:initialize-localization}).

The main part of the algorithm is a worklist loop
(lines~\ref{line:localize-loop-start} to \ref{line:clear-f-}). In this loop, the
algorithm first processes all unwanted but appearing faults ($F^+$,
line~\ref{line:repair-t+}) by computing the repair of $F^+$. The result is a
subset of tuples in the diff such that the faults $F^+$ no longer appear when
the subset is excluded from the diff. In the provenance system, negations are
treated as EDB tuples, and thus the resulting repair may contain negated tuples.
These negated tuples are added to $F^-$ (line~\ref{line:add-!t-to-F-}) since a
tuple appearing in $F^+$ may be caused by a negated tuple disappearing. The
algorithm then repairs the tuples in $F^-$ by computing the dual problem, i.e.,
localizing $F^-$ with respect to $\Delta E_{2 \rightarrow 1}$. Again, this
process may result in negated tuples, which are added to $F^+$, and the loop
begins again. This worklist loop must terminate, due to the semantics of
stratified negation, as discussed above. At the end of the worklist loop, $L$
contains a candidate repair.

While Alg.~\ref{alg:full-fault-repair} presents a full algorithm for rollback,
the fault localization problem can be solved similarly. Since rollback and
localization are dual problems, the full fault localization algorithm swaps
Rollback-Repair in line~\ref{line:repair-t+} and Localize-Faults in
line~\ref{line:localize-t-}.

\paragraph{Example} We demonstrate how our algorithms work by using our running
example. Recall that we introduce an incremental update consisting of inserting
two tuples: 

\texttt{assign(upgradedSession, userSession)} and
\texttt{load(userSession, admin, session)}. As a result, the system computes the
unwanted fault tuple \texttt{alias(userSession, sec)}. To rollback the
appearance of the fault tuple, the algorithms start by computing its provenance,
as shown in Figure~\ref{fig:inc-proof-tree}. The algorithm then creates a set of
ILP constraints, where each tuple (with shortened variables) represents an ILP
variable:
\begin{align*}
    &\text{maximize} \sum \mathtt{load(u,a,s)} \text{ such that} \\
    &\mathtt{load(u,a,s)} - \mathtt{vpt(u,L2)} \leq 0,\\
    &\mathtt{vpt(u,L2)} - \mathtt{alias(u,s)} \leq 0,\\
    &\mathtt{alias(u,s)} = 0
\end{align*}
For this simple ILP, the result indicates that the insertion of
\texttt{load(userSession, admin, session)} should be rolled back to fix the
fault.

\subsection{Correctness and Optimality}

In this section, we discuss the correctness and optimality of our algorithms.
Consider the problem set up with a Datalog program $P$, an EDB $E_1$, an incremental
update diff $\Delta E_{1 \rightarrow 2}$, and a set of fault tuples $F$.
For the fault localization algorithm, correctness implies that the result
reproduces the faults, i.e., that $F \subseteq (E_1 \uplus \delta E)$.
Meanwhile, the correctness of a rollback repair implies that the
result prevents faults from appearing, i.e., that
$(E_1 \uplus (\Delta E_{1 \rightarrow 2} \setminus \delta_{\times} E)) \cap F = \emptyset$

Optimality is measured by how minimal the solution is. For both fault localization
and rollback repair, a solution is optimal if there is no smaller subset of
the solution which correctly solves the problem.



\paragraph{Fault Localization.} The correctness of fault localization (Algorithm~\ref{alg:localize-fault}) lies in
the semantics of the proof trees. Consider a single proof tree. If every EDB tuple
in the proof tree exists as input, then the resulting tuple at the root of the
proof tree would be computed by the Datalog program. Therefore, since the fault
localization algorithm returns \emph{all} $\Delta EDB$ tuples in the proof tree,
then the resulting fault tuple will be in the result.

The optimality of the fault localization result is dependent on the
properties of the proof trees produced in the Datalog engine. If these proof trees
are minimal in terms of the number of EDB tuples, then the fault localization
result will also be minimal in size. This property can be guaranteed depending
on how the proof tree generation is implemented (see \cite{zhao2020debugging}),
however the details are outside the scope of this paper.


\paragraph{Rollback Suggestions.} The crucial step in the rollback repair algorithm
(Algorithm~\ref{alg:repair-fault}) involves encoding the proof trees as an ILP.
These ILP constraints directly encode the logical formulae representing the semantics
of the proof trees, with additional constraints asserting that the faulty tuples
must be false (i.e., that they are not computed by the EDB satisfying the
ILP). Therefore, the correctness of the rollback repair algorithm results
from the correctness of the ILP encoding and solving.

To consider the optimality of a rollback suggestion, we first note that the algorithm
uses \emph{all} non-cyclic proof trees. This means that the properties of each
proof tree do not affect optimality, but rather, the optimality is a result of
the maximization constraint in the ILP encoding. This constraint represents
that the maximum number of tuples in $\Delta EDB$ must be kept in the solution,
which is equivalent to saying that the rollback repair includes the \emph{minimum}
number of necessary tuples. Hence, the solution is indeed optimal.



\paragraph{Full Algorithm.} Each component of the full algorithm is correct, as
discussed above, and therefore it only remains to be shown that considering the
dual problem for negations is correct. This correctness is discussed in
Section~\ref{sec:algo-extensions}, and thus the full algorithm identifies a
correct debugging suggestion.

However, the full algorithm is not necessarily optimal in the presence of
negation. For example, consider when an initial debugging suggestion includes a
negated tuple, $t$. Then, the full algorithm computes the dual problem of
localizing the appearance of $t$ with the opposite diff $\Delta E_{2 \rightarrow
1}$. However, this opposite diff does not consider the initial debugging
suggestion (only the negated tuple), and thus, the result may not be optimal. In
practice, this sub-optimality rarely affects the solution, and the result is
generally optimal or close to optimal.

\paragraph{Complexity.} The fault localization algorithm
(Algorithm~\ref{alg:localize-fault}) simply computes the provenance for each
fault tuple. From \cite{zhao2020debugging}, computing a proof tree requires $O(h
\log^m \vert \IDB \vert)$ time, where $h$ is the height of the proof tree, $m$ is the nesting
depth of the joins in the Datalog program, and $\vert \IDB \vert$ is the total number of tuples
computed in the IDB.

The rollback suggestion algorithm computes the full provenance of the fault
tuple, which requires up to $n$ applications of the provenance algorithm, if
there are $n$ proof trees for the tuple.
However, the integer linear programming portion is exponential in complexity,
with branch-and-bound-based algorithms~\cite{clausen1999branch} taking
$O(2^{\vert V \vert})$ runtime, where $\vert V \vert$ is the number of
variables. In our case, there is one variable for each tuple in the full
provenance, which is up to $\vert \IDB \vert$ in the worst case. This dominates
the runtime of our algorithm, resulting in a total complexity of $O(nh \log^m
\vert \IDB \vert) + O(2^{\vert \IDB \vert})$.

In practice, however, the size of the full provenance for a fault tuple is far
smaller than the full IDB, resulting in reasonable real-world performance even
for large Datalog programs.

For comparison, the delta debugging approach only needs to check a linear number
of subsets of the incremental update. However, for each subset, delta debugging
needs to evaluate an (incremental) Datalog program, which is polynomial in
complexity, but sometimes prohibitive in practice.


%% file: src/impl.tex
\section{Implementation}
\label{sec:implementation}

The implementation of our algorithms first involved extending the Souffl\'e
Datalog engine~\cite{jordan2016souffle}. Souffl\'e already includes utilities
for computing proof trees (also called provenance)~\cite{zhao2020debugging} and
incremental evaluation~\cite{zhao2021towards}.
To support fault localization and rollback suggestions, we extended
Souffl\'e to support \emph{incremental provenance}. This involved interoperability
between the provenance and incremental evaluation portions, to allow the
provenance mechanism to use the same instrumentation originally designed
for incremental evaluation.

We implemented the fault
localization and repair algorithms using Python\footnote{Available at
\url{github.com/davidwzhao/souffle-fault-localization}}. The implementations
of the Fault Localization and Rollback Repair algorithms follow directly
from their presentations in this paper. For any operations which require
calling a Datalog engine, we call out to our modified Souffl\'e engine.
One potential source of inefficiency in our implementation is that Souffl\'e
does not have direct Python interoperability, so we have to read/write
tuples through the filesystem or pipes to interact with Souffl\'e. For solving
integer linear programs, we use the GLPK solver~\cite{glpk} in the PuLP Python library.

For the full algorithm, we need to compute the dual versions of each problem.
For efficiency, we do not construct the full dual version of the problem as
they are needed, but instead, we maintain two instances of Souffl\'e: one
for the \emph{forward} problem, and one for the \emph{reverse} direction.
Using these two instances of Souffl\'e, we can easily compute the fault
localizations or rollback suggestions as needed, without re-instantiating
the Datalog engine.


%% file: src/exp.tex
\section{Experiments}
\label{sec:experiment}

This section evaluates our technique on real-world benchmarks to determine
its effectiveness and applicability. We consider the following research
questions:

\begin{itemize}
    \item \unskip RQ1: Is the new technique faster than a delta-debugging
        strategy?
    \item \unskip RQ2: Does the new technique produce more precise
        localization/repair candidates than delta debugging?
\end{itemize}

\begin{table}
    \caption{Repair size and runtime of our technique compared to delta debugging}
    \label{tbl:results}
    \footnotesize
    \centering
    \begin{tabular}{lc||r|r|r|r||r|r||r}
        \hline
        & & \multicolumn{4}{c||}{Rollback Repair} & \multicolumn{2}{c||}{Delta Debugging} & \\
        \cline{3-6} \cline{7-8}
        Program & No. & Size & Overall (s) & Local(s) & Repair(s) & Size & Runtime (s) & Speedup \\
        \hline
        antlr & 1 & 2 & 73.6 & 0.51 & 73.1 & 3 & 3057.8 & 41.5 \\
         & 2 & 1 & 79.4 & 0.00 & 79.4 & 1 & 596.5 & 7.5 \\
         & 3 & 1 & 0.95 & 0.95 & - & 1 & 530.8 & 558.7 \\
         & 4 & 2 & 77.8 & 1.89 & 75.9 & 3 & 3017.6 & 38.8 \\
        bloat & 1 & 2 & 3309.5 & 0.02 & 3294.1 & 2 & 2858.6 & 0.9 \\
         & 2 & 1 & 356.3 & 0.00 & 355.4 & 1 & 513.6 & 1.4 \\
         & 3 & 1 & 0.33 & 0.33 & - & 1 & 557.7 & 1690.0 \\
         & 4 & 3 & 3870.6 & 0.10 & 3854.7 & 2 & 2808.3 & 0.7 \\
        chart & 1 & 1 & 192.6 & 0.00 & 192.6 & 1 & 685.0 & 3.6 \\
         & 2 & 1 & 3.01 & 3.01 & - & 1 & 675.3 & 224.4 \\
         & 3 & 1 & 78.8 & 0.00 & 78.8 & 1 & 667.6 & 8.5 \\
         & 4 & 2 & 79.9 & 3.24 & 76.7 & 3 & 3001.1 & 37.6 \\
        eclipse & 1 & 2 & 177.3 & 0.04 & 177.2 & 3 & 2591.2 & 14.6 \\
         & 2 & 1 & 79.2 & 0.00 & 79.1 & 1 & 416.1 & 5.3 \\
         & 3 & 1 & 0.12 & 0.12 & - & 1 & 506.3 & 4219.2 \\
         & 4 & 3 & 91.9 & 0.09 & 91.8 & 3 & 2424.4 & 26.4 \\
        fop & 1 & 2 & 83.8 & 0.05 & 83.8 & 2 & 3446.6 & 41.1 \\
         & 2 & 1 & 76.9 & 0.00 & 76.9 & 1 & 670.7 & 8.7 \\
         & 3 & 1 & 0.66 & 0.66 & - & 1 & 721.8 & 1093.6 \\
         & 4 & 6 & 74.8 & 0.50 & 74.3 & \multicolumn{2}{c||}{Timeout (7200)} & 96.3+ \\
        hsqldb & 1 & 2 & 83.3 & 0.04 & 83.3 & 2 & 2979.2 & 35.8 \\
         & 2 & 1 & 79.4 & 0.00 & 79.4 & 1 & 433.8 & 5.5 \\
         & 3 & 1 & 74.0 & 0.00 & 74.0 & 1 & 663.1 & 9.0 \\
         & 4 & 3 & 75.5 & 0.04 & 75.5 & 5 & 6134.8 & 81.3 \\
        jython & 1 & 1 & 83.3 & 0.00 & 83.3 & 1 & 609.4 & 7.3 \\
         & 2 & 1 & 78.2 & 0.00 & 78.2 & 1 & 590.4 & 7.5 \\
         & 3 & 1 & 76.6 & 0.00 & 76.6 & 1 & 596.2 & 7.8 \\
         & 4 & 1 & 75.8 & 0.00 & 75.8 & 1 & 587.6 & 7.8 \\
        luindex & 1 & 2 & 81.3 & 0.07 & 81.2 & 3 & 2392.1 & 29.4 \\
         & 2 & 1 & 79.8 & 0.00 & 79.8 & 1 & 511.0 & 6.4 \\
         & 3 & 1 & 0.10 & 0.10 & - & 1 & 464.8 & 4648.0 \\
         & 4 & 4 & 77.9 & 0.12 & 77.8 & 5 & 4570.4 & 58.7 \\
        lusearch & 1 & 2 & 110.2 & 0.06 & 110.0 & 3 & 2558.8 & 23.2 \\
         & 2 & 1 & 1062.1 & 0.00 & 1057.4 & 1 & 370.4 & 0.3 \\
         & 3 & 1 & 0.12 & 0.12 & - & 1 & 369.6 & 3080.0 \\
         & 4 & 2 & 294.2 & 0.06 & 293.2 & 3 & 2420.9 & 8.2 \\
        pmd & 1 & 2 & 78.1 & 0.02 & 78.1 & 3 & 3069.8 & 39.3 \\
         & 2 & 1 & 77.0 & 0.00 & 77.0 & 1 & 600.2 & 7.8 \\
         & 3 & 1 & 0.08 & 0.08 & - & 1 & 717.8 & 8972.5 \\
         & 4 & 3 & 74.3 & 0.08 & 74.2 & 3 & 2828.3 & 38.1 \\
        xalan & 1 & 1 & 84.9 & 0.00 & 84.9 & 1 & 745.3 & 8.8 \\
         & 2 & 1 & 82.2 & 0.00 & 82.2 & 1 & 728.9 & 8.9 \\
         & 3 & 1 & 100.1 & 0.00 & 100.1 & 1 & 1243.7 & 12.4 \\
         & 4 & 1 & 521.6 & 0.00 & 518.3 & 1 & 712.5 & 1.4 \\
        \hline
    \end{tabular}
\end{table}

\paragraph{\textbf{Experimental Setup:}}\footnote{We use an Intel Xeon Gold 6130
with 192 GB RAM, GCC 10.3.1, and Python 3.8.10}

Our main point of comparison in our experimental evaluation is the delta
debugging approach, such as that used in the ProSynth Datalog synthesis
framework~\cite{raghothaman2019provenance}. We adapted the implementation of
delta debugging used in ProSynth to support input tuple updates. Like our fault
repair implementation, the delta debugging algorithm was implemented in Python;
however, it calls out to the standard Souffl\'e engine since that provides a
lower overhead than the incremental or provenance versions.

For our benchmarks, we use the Doop program analysis framework~\cite{doop09}
with the DaCapo set of Java benchmarks~\cite{blackburn2006dacapo}. The analysis
contains approx. 300 relations, 850 rules, and generates approx. 25 million
tuples from an input size of 4-9 million tuples per DaCapo benchmark. For each
of the DaCapo benchmarks, we selected an incremental update containing 50 tuples
to insert and 50 tuples to delete, which is representative of the size of a
typical commit in the underlying source code. From the resulting IDB changes, we
selected four different arbitrary fault sets for each benchmark, which
may represent an analysis error. 

\paragraph{\textbf{Performance:}} The results of our experiments are shown in
Table~\ref{tbl:results}. Our fault repair technique performs much better overall
compared to the delta debugging technique. We observe a geometric mean speedup
of over 26.9$\times$ \footnote{We say ``over'' because we bound timeouts to 7200
seconds.} compared to delta debugging. For delta debugging, the main cause of
performance slowdown is that it is a black-box search technique, and it requires
multiple iterations of invoking Souffl\'e (between 6 and 19 invocations for the
presented benchmarks) to direct the search. This also means that any
intermediate results generated in a previous Souffl\'e run are discarded since
no state is kept to allow the reuse of results. Each invocation of Souffl\'e
takes between 30-50 seconds, depending on the benchmark and EDB. Thus, the
overall runtime for delta debugging is in the hundreds of seconds at a minimum.
Indeed, we observe that delta debugging takes between 370 and 6135 seconds on
our benchmarks, with one instance timing out after two hours (7200 seconds).

On the other hand, our rollback repair technique calls for provenance
information from an already initialized instance of incremental Souffl\'e. This
incrementality allows our technique to reuse the already computed IDB for each
provenance query. For eight of the benchmarks, the faults only contained missing
tuples. Therefore, only the Localize-Faults method was called, which only
computes one proof tree for each fault tuple and does not require any ILP
solving. The remainder of the benchmarks called the Rollback-Repair method,
where the main bottleneck is for constructing and solving the ILP instance. For
three of the benchmarks, \texttt{bloat-1}, \texttt{bloat-4}, and
\texttt{lusearch-2}, the runtime was slower than delta debugging. This result is
due to the fault tuples in these benchmarks having many different proof trees,
which took longer to compute. In addition, this multitude of proof trees causes
a larger ILP instance to be constructed, which took longer to solve.

\paragraph{\textbf{Quality:}} While the delta debugging technique produces
1-minimal results, we observe that despite no overall optimality guarantees, the
results show that our approach was able to produce more minimal repairs in 27\%
of the benchmarks. Moreover, our rollback repair technique produced a larger
repair in only one of the benchmarks. This difference in quality is due to the
choices made during delta debugging. Since delta debugging has no view of the
internals of Datalog execution, it can only partition the EDB tuples randomly.
Then, the choices made by delta debugging may lead to a locally minimal result
that is not globally optimal. For our fault localization technique, most of the
benchmarks computed one iteration of rollback repair and did not encounter any
negations. Therefore, due to the ILP formulation, the results were optimal in
these situations. In one case, the rollback repair encountered a negation and
flipped to the dual fault localization problem, resulting in the suboptimality.
Despite our technique overall not being theoretically optimal, it still produces
high-quality results in practice.

%% file: src/related.tex
\section{Related Work}
\label{sec:related-work}

\paragraph{Logic Programming Input Repair.} A plethora of logic programming
paradigms exist that can express diagnosis and repair by EDB
regeneration~\cite{alp1,dlv,asp,cav17,LiuMSR23}. Unlike these logic programming
paradigms, our technique is designed to be embedded in high-performance modern
Datalog engines. Moreover, our approach can previous
computations (proof trees and incremental updates) to localize and repair only
needed tuples. This bounds the set of repair candidates and results in apparent
speedups. Other approaches, such as the ABC Repair System~\cite{li2018abc}, use
a combination of provenance-like structures and user-guided search to localize
and repair faults. However, that approach is targeted at the level of the
Datalog specification and does not always produce effective repairs. Techniques
such as delta debugging have recently been used to perform state-of-the-art
synthesis of Datalog programs efficiently~\cite{raghothaman2019provenance}. Our
delta debugging implementation adapts this method, given it produces very
competitive synthesis performance and can be easily re-targeted to diagnose and
repair inputs.

\paragraph{Database Repair.} Repairing inconsistent databases with respect to
integrity constraints has been extensively investigated in the database
community~\cite{dbrepair1,dbrepair2,dbrepair3,dbrepair4}. Unlike our approach,
integrity constraints are much less expressive than Datalog; in particular, they
do not allow fixpoints in their logic. The technique in \cite{dbrepair4} shares
another similarity in that it also presents repair for incremental SQL
evaluation. However, this is limited to relational algebra, i.e., SQL and
Constrained Functional Dependencies (CFDs) that are less expressive than
Datalog. A more related variant of database repair is consistent query answering
(CQA)\cite{dbrepair2,dbrepair3}. These techniques repair query answers given a
database, integrity constraints and an SQL query. Similarly, these approaches do
not support recursive queries, as can be expressed by Datalog rules.

\paragraph{Program Slicing.} Program
slicing~\cite{weiser1984program,binkley1996program,ezekiel2021locating,harman2001overview}
encompasses several techniques that aim to compute portions (or
\emph{slices}) of a program that contribute to a particular output result. For
fault localization and debugging, program slicing can be used to localize slices
of programs that lead to a fault or error. The two main approaches are
\emph{static} program slicing, which operates on a static control flow graph,
and \emph{dynamic} program slicing, which considers the values of variables or
execution flow of a particular execution. As highlighted by~\cite{Cheney07},
data provenance is closely related to slicing. Therefore, our technique can be
seen as a form of static slicing of the Datalog EDB with an additional rollback
repair stage.

\paragraph{Database Rollback.}
Database transaction rollback and partial rollback are well established~\cite{dbrollback,dbrollback2} and supported in many DBMS's~\cite{oraclerollback,msrollback}. These techniques often 
perform rollback for a transaction in the context of data recovery, by logging the effects of each action 
in the transaction. Techniques such as~\cite{AntonopoulosBCD19} improve rollback time 
by using versioning information. These techniques are limited to SQL transactions while our technique targets recursive datalog queries in an incremental update setting. For the static analysis use case, to the best of our knowledge, we are the first to provide an automated partial commit rollback mechanism based on the analysis output. Nevertheless, it is interesting future work to investigate 
if our technique can assist in making more efficient partial rollbacks in a DBMS setting.

\paragraph{Automated Commit Rollback.}
There is not a lot of work in the literature on automatically detecting and partially rolling back 
buggy commits, despite several studies~\cite{st1,idcommit} highlighting the
benefits of identifying such commits and rolling them back as soon as
possible. The closest works to ours are techniques~\cite{idcommit,idcommit2,idcommit3,idcommit4} that seek to identify
through statistical models commits that are most likely to be reverted. In contrast, our technique works 
with a static analyzer that detects bugs in code, and provides users with the option to 
partially revert the commit so the bug is eliminated. 

%% file: src/concl.tex
\section{Conclusion}
We have presented a new debugging technique that localizes faults and provides
rollback suggestions for Datalog program inputs. Unlike previous approaches, our
technique does not entirely rely on a black-box solver to perform the underlying
repair. Instead, we utilize incremental provenance information. As a result, our
technique exhibits speedups of 26.9$\times$ compared to delta debugging and
finds more minimal repairs 27\% of the time.

There are also several potential future directions for this research. One
direction is to adopt these techniques for different domain areas outside the
use cases of program analyses.